\documentclass{vietnam}

\def\Journal#1#2#3#4{{#1} {\bf #2}, #3 (#4)}

\def\PRD{{\em Phys. Rev.} D}

\def\be{\begin{equation}}
\def\ee{\end{equation}}
\def\bea{\begin{eqnarray}}
\def\eea{\end{eqnarray}}

\newcommand{\Photo}{}

\begin{document}
\vspace*{4cm}
\title{NEWS : a new spherical gas detector for very low mass WIMP detection}

%\author{ G. Gerbier, for the Saclay team of the  -{\bf N}ew  {\bf E}xperiments  {\bf W}ith  {\bf S}phere- network }
\author{ G. Gerbier, I. Giomataris, P. Magnier, A. Dastgheibi, M. Gros, D. Jourde, E. Bougamont, X.F. Navick, T. Papaevangelou, J. Galan,  J. Derre, I. Savvidis, G. Tsiledakis.  Saclay team member of  the -{\bf N}ew  {\bf E}xperiments  {\bf W}ith  {\bf S}phere- network }

\address{IRFU, CEA Saclay, Gif s Yvette, France}

\maketitle\abstracts{
The main characteristics of a new concept of spherical gaseous detectors, with some details on its operation are first given. The very low energy threshold of such detector has led to investigations of its potential performance for dark matter particle searches, in particular low mass WIMP's : original methods for energy and fiducial volume calibration and background rejection are described and preliminary results obtained  with a low radioactivity prototype operated in Laboratoire Souterrain de Modane ("Frejus" lab) are presented. Typical expected sensitivities in cross section for low mass WIMP's are also shown, and other applications briefly discussed.
}

\section{Introduction}

Dark matter is now clearly an essential ingredient of our understanding of the Universe. Its nature is still unknown but -massive- neutral particles, non relativistic at decoupling time, are a generic class of  well motivated candidates. Recent -non-findings at LHC indicate that the preferred candidate,  the LSP or neutralino, has somewhat lost its credibility as the main constituent of dark matter~\cite{lhc}. While not all models of SUSY have not yet been rejected, the MSSM is strongly  disfavored. Furthermore, the absence of any hint of new physics leaves the dark matter hunters in front of a wide open space to explore.

In particular the somewhat rather well defined mass range from typically few 10ths to thousands of GeV is to be extended. Several new approaches (dark sector, asymmetric dark matter, U-boson, generalized effective theory ...) open the way to less paradigmatic candidates, with lower mass and/or more complex couplings that the traditional spin (in)dependant ones~\cite{newimp}. 

These conclusions happen to come at the same time as some  febrility on experimental side where unexplained excesses of events interpreted as hints of signals popped up recently, however all are close to the energy threshold of detectors~\cite{ggrev}.

Indeed, many new experimental ideas come to maturity. The presently studied detector, a spherical gaseous detector, initially proposed  by I Giomataris~\cite{sph}, will allow to explore new parameter space for dark matter. A main goal will be to search for very-light dark matter particles with mass much lower than 10 GeV and shed light on the region of 7 GeV,  where some claims are advocated.

\section{Principle and main features of  the spherical gas detector}

The studied detectors consist of a spherical metallic vessel -from 0.15 m to 1.3 m in diameter- and a small metallic ball from 3 to 14 mm in diameter located at the center of the vessel. The ball is maintained in the center of the sphere by a rod and is set at high voltage. The field, varying as 1/r$^2$ is highly inhomogeneous along the radius, allowing the electrons to drift to the central sensor in low field regions constituting most of the volume, while they trigger an avalanche within few mm around the sensor (Figure 1a). The amplification capability combined with the very low capacitance of the sensor allows to reach easily sub-kev threshold, and, in particular settings, single ionization electron sensitivity. It should be noted that the threshold does not depend on the size of the vessel, anticipating the possibility to handle rather large mass of targets read by a single channel.

\begin{figure}
\begin{minipage}{0.5\linewidth}
\centerline{\includegraphics[width=0.9\linewidth]{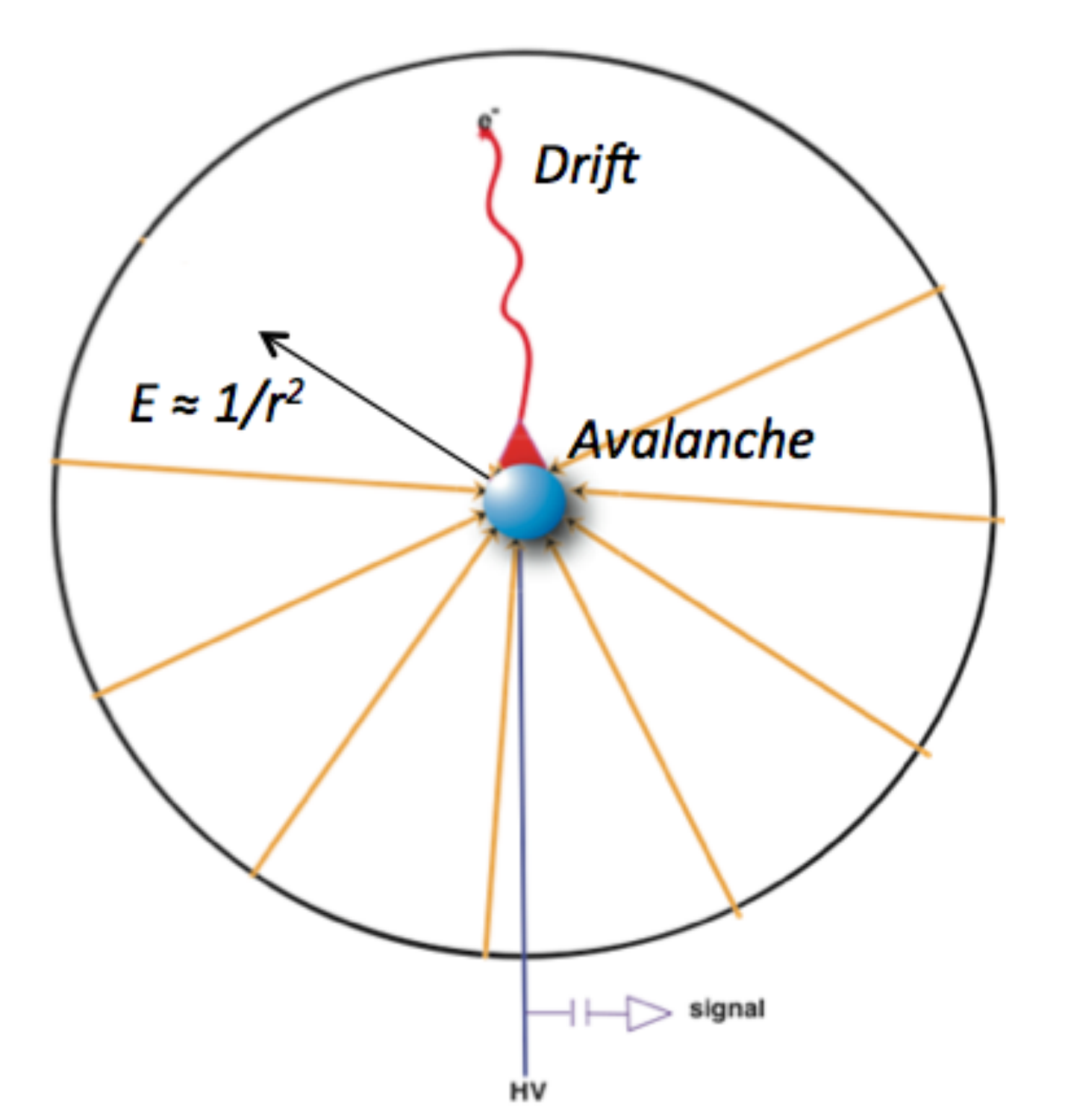}}
\end{minipage}
\begin{minipage}{0.5\linewidth}
\centerline{\includegraphics[width=0.9\linewidth]{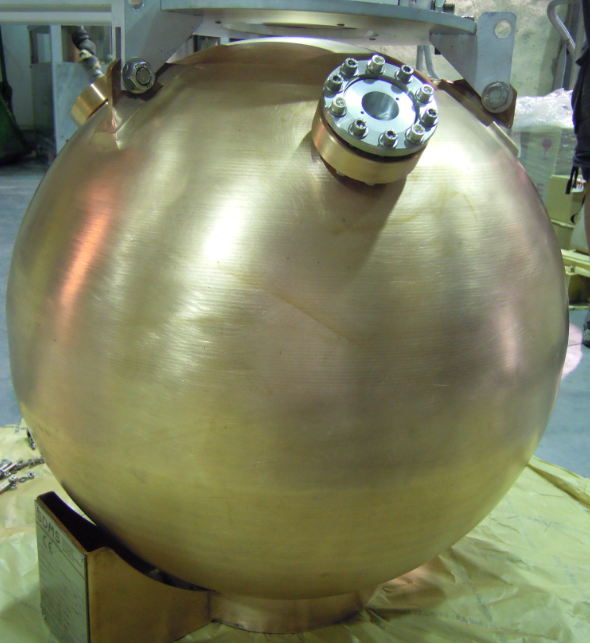}}
\end{minipage}
\caption{a left) Principle of spherical gas detector - b right ) 60cm prototype made of low activity Copper}
\label{fig:radish}
\end{figure}

Other key advantages of this detector are its fiducialisation capability and the possibility to isolate point like energy deposition by pulse shape analysis.

Last but not least, the possibility to vary the pressure, the nature of the gas, (from Helium to Xenon and possibly Hydrogen) allows to cover a wide range of WIMP mass, and provide many knobs for redundancy to check the possible presence of a signal.

Getting a low threshold for large mass however requires to keep sizable gains at high pressure, which leads to the choice of a small diameter for the sensor. However, the field which is roughly proportional to the diameter of the sensor becomes very low at large radii and may not be strong enough to drift electrons to the sensor.  Studies on sensor are pursued to decouple these two requirements.

Calibrations and preliminary data obtained with a low activity prototype installed in the underground laboratory of Modane  (Figure1b) are presented in next sections.

\section{Experimental details on detectors and set up}

Calibrations and tests have been performed in a surface lab at  Saclay  with a spherical copper vessel of 130 cm in diameter, used up to 1 bar pressure, and with a low activity prototype of 60 cm in diameter, installed within lead and polyethylene shields  in Laboratoire Souterrain de Modane, designed to hold up to 10 bars pressure.

The detectors were operated with various spherical sensors, ranging from 3 to 16 mm in diameter, hold, through adequate isolator, by a metallic rod at ground. 
Two types of sensors were studied : one with an "umbrella", mounted on the rod a few cm from the spherical sensor, and set to an intermediate high voltage, the other with no "umbrella".
Set-ups were tuned thanks to electrostatic simulations to obtain the best homogeneity of the field inside the vessel.
Anyhow, the homogeneity of response is ultimately assessed by the symmetry of the peak obtained  with mono-energetic photons from radioactive sources converted homogeneously inside the volume of gas.  

Care was taken to have good vacuum and low leak rate, typically less than 10$^{-7}$ mb/s. Applied voltages ranged from 1000 V to 6000 V, depending on pressure, from few mb to 4 bars.

The signal is extracted from the HV wire through a capacitor, amplified by a charge amplifier, with time constants ranging from 30 $\mu$s to 300 $\mu$s, digitized at frequency from 0.5 to 4 MHz, and sent to a computer which performs a software trigger after adequate noise filtering to obtain the lowest threshold amplitude.

To decrease the energy threshold or to operate the detector at lower gain, our current preamplifier (600 electron RMS noise) will be soon replaced by a very-low noise one ($< $70 electron RMS). 

\begin{figure}
\begin{minipage}{0.5\linewidth}
\centerline{\includegraphics[width=0.9\linewidth]{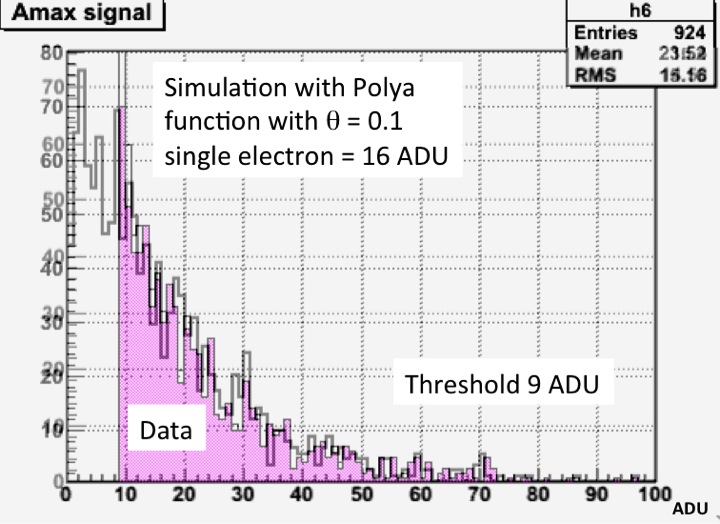}}
\end{minipage}
\begin{minipage}{0.5\linewidth}
\centerline{\includegraphics[width=0.9\linewidth]{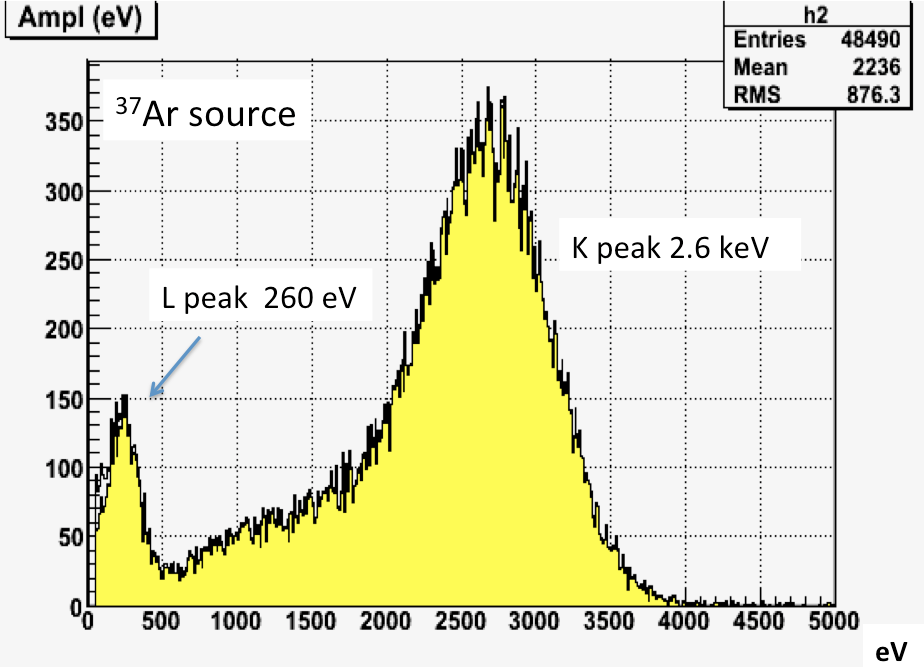}}
\end{minipage}
\caption{a left) Single electron response, note the 17 eV threshold, see text - b right) Low energy calibration with $\rm ^{37}$Ar, note the 50 eV threshold.}
\label{fig:radish}
\end{figure}

\section{Calibrations}

\subsection{Single electron response }

A dedicated study was performed with a pulsed UV lamp, shining through a MgF${_2}$ window. The UV's hitting the internal surface of the cavity extract electrons from Cu atoms, which then drift to the sensor. The timing of the pulsed UV lamp  allows to measure the drift times. By inserting optical attenuators along the path of the UV's, the intensity of each pulse can be tuned to induce single electron production on the surface.

The figure 2a shows the spectrum of amplitudes obtained with Ar/CH${_4}$ (98/2 \% in volume) mix at a pressure of 250 mb, under the conditions of single electron production. Superimposed is a simulation of pulses with a polya distribution with parameter $\theta$ of 0.1. 
Given the w parameter of Argon of 30 eV, the obtained threshold was equivalent to an energy threshold of 17 eV.

As anticipated, given the value of $\theta$ parameter, the resolution of a single electron is quite poor. However, the knowledge of this parameter is important to simulate the accurate amplitude spectrum expected in case of very small  numbers of produced electrons.

\subsection{Ar37 calibration and homogeneity of response}

We obtained a  volume calibration with low energy photons by using the $\rm ^{37}$Ar radioactive isotope. $\rm ^{37}$Ar decays through electron capture with a period of 35 days, and gives rise to isolated K Xray of 2.6 keV and L Xray of 260 eV from $\rm ^{37}$Cl.  $\rm ^{37}$Ar was obtained through (n,$\alpha$) reaction on $\rm ^{40}$Ca. A low intensity source (few Hz) has then been produced by irradiating few hundreds of grams of Calcium salt powder to a source of $\simeq$ 10$^{7}$ fast n/s  during few days. 

The figure 2b shows the spectrum obtained with this method applied to the low activity prototype in LSM. The 260 eV peak can be clearly seen, together with a reasonably symmetric gaussian peak for the 2.6 keV K Xray. We have hints that the left over asymmetry may be due to degassing from powder which degrades the quality of gas and electron collection at the periphery of the sphere, under very low drift fields.

\subsection{Neutron and volume calibration}

As underlined in the introduction, fiducialisation can be obtained by measuring the rise time of the pulse. For point like energy deposition, the larger is the radius, the larger is the longitudinal diffusion, the larger is the rise time.
To obtain a clean sample of  point like energy deposition interactions, we used a Am-Be neutron source to induce recoils from the nuclei in the gas. We used a mixture of Neon/Helium/CH${_4}$ (49/49/2 \%)  at 2 bars allowing to obtain a wide range of elastic nuclear recoils energies. Figure 3a shows the scatter plot of rise time vs energy of events acquired with a neutron source (green dots) for 24 h and with no source (blue dots) for 72h, under the same detector conditions. For the neutron source run, only events with a shape matching the one expected from a local energy deposition are plotted.  The green dots above roughly 80 keV correspond to nuclear recoils of He, extending up to 500 keV electron equivalent energy, while the ones at lower energy correspond mostly to  Neon recoils. Note that the green dot population is confined in the range 10-30 $\mu$s defining then the volume interactions. Note also that the events of the blue population below 100 keV are mostly above a rise time of 30 $\mu$s.

\begin{figure}
\begin{minipage}{0.5\linewidth}
\centerline{\includegraphics[width=0.9\linewidth]{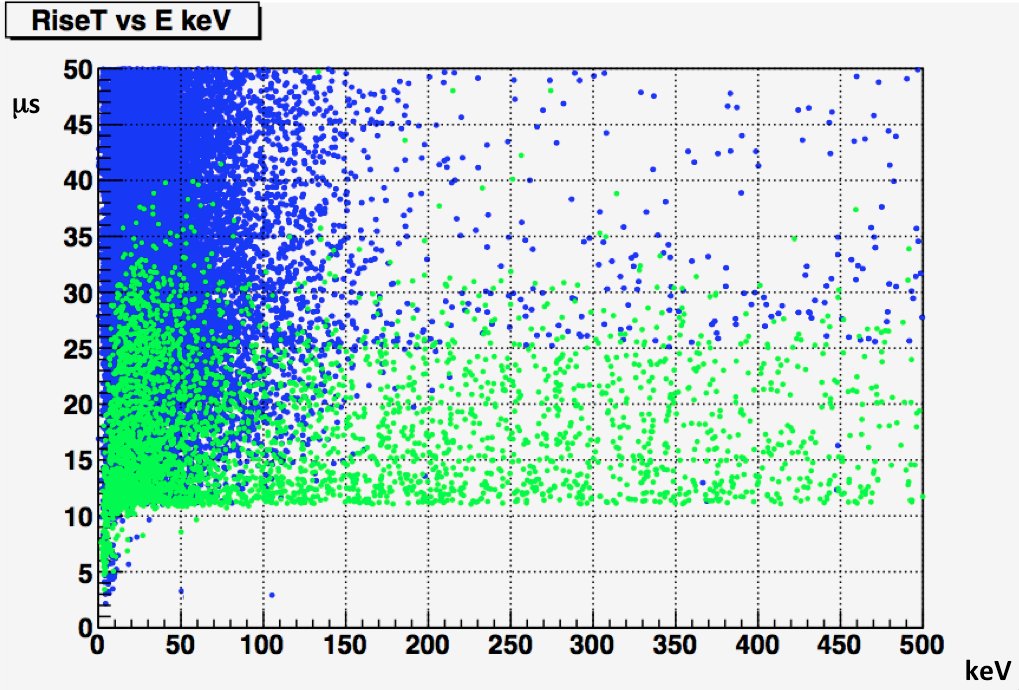}}
\end{minipage}
\begin{minipage}{0.5\linewidth}
\centerline{\includegraphics[width=0.9\linewidth]{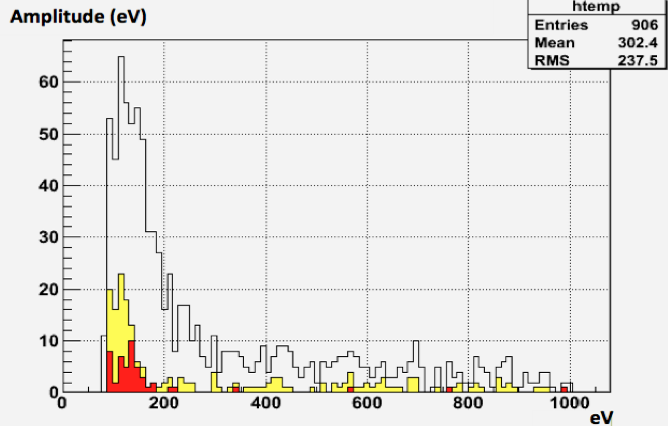}}
\end{minipage}
\caption{a left) Risetime vs energy diagrams for neutron calibration and background data, see text - b right ) Low energy spectrum of background data - see text.}
\label{fig:radish}
\end{figure}

\section{Data obtained in underground lab}

Data have been obtained with the SEDINE prototype installed within lead (10 cm) and polyethylene (30 cm) shields.
Radon free air ($< $0.1 Bq/m$^{3}$) provided by the LSM facility is flushed inside the volume between the sphere and the lead shield when taking background data. Figure 3b shows the energy distribution of a short no source run at low energy. The full line shows the total rate, without cut, the yellow histogram the events kept in the rise time window corresponding to the full volume, the red histogram are the events kept in a rise time window adjusted for background rejection and after selection of the point like pulse shapes. The corresponding rate at 200 eV corresponds to around 100 events per kev.kg.d. Given that various measurements show that the present background is dominated by surface $^{210}$Pb contamination inside and outside of the vessel, between 10 and 100 times the expected one, this is a quite promising situation. We recently have performed a new chemical attack to remove $^{210}$Pb and added a new Cu shield between the sphere and the lead. Measurements are ongoing.

\begin{figure}
\begin{minipage}{0.5\linewidth}
\centerline{\includegraphics[width=0.9\linewidth]{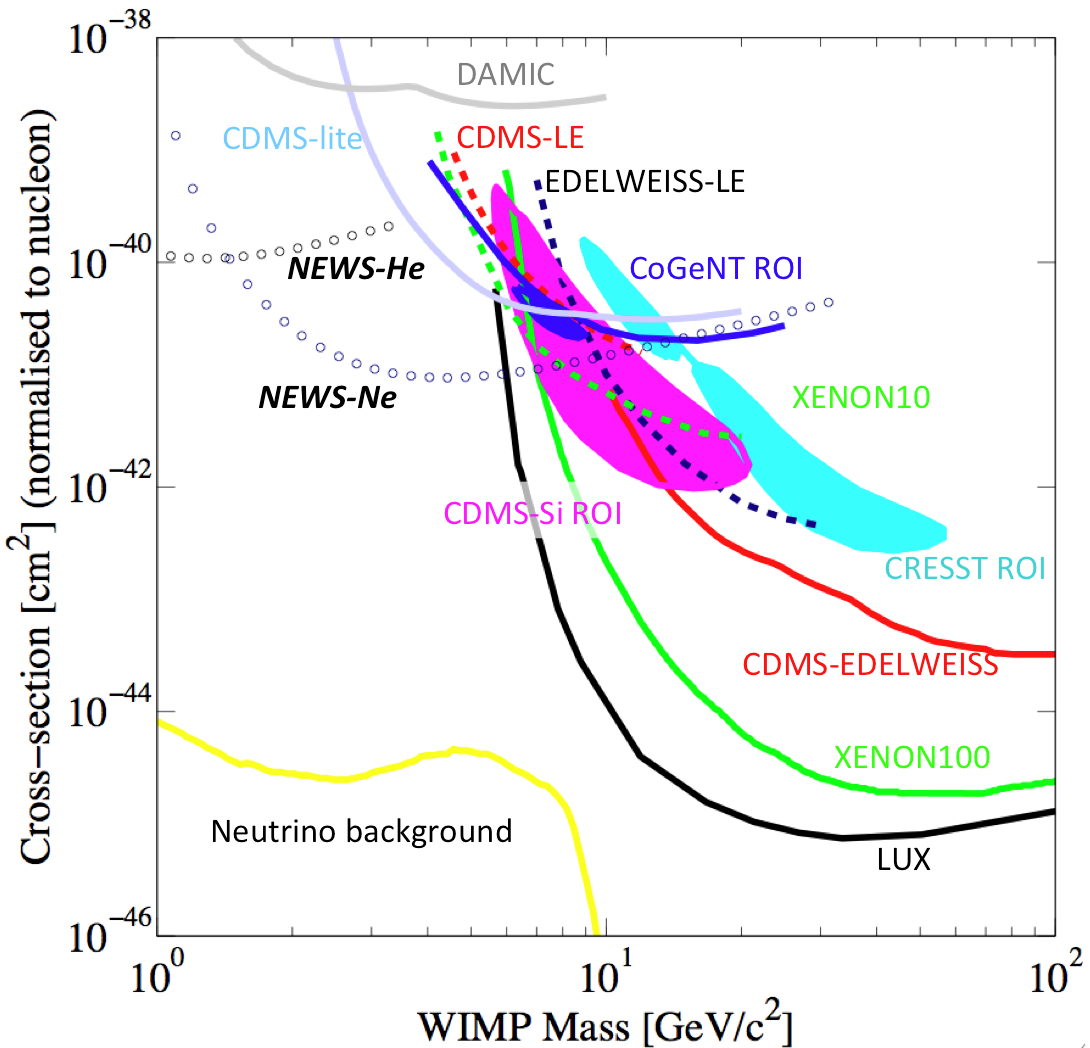}}
\end{minipage}
\begin{minipage}{0.5\linewidth}
\centerline{\includegraphics[width=0.9\linewidth]{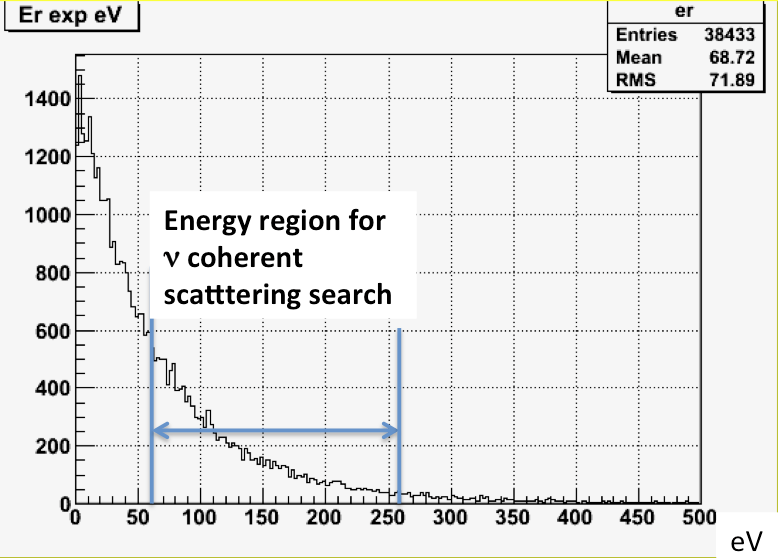}}
\end{minipage}
\caption{a left) Anticipated sensitivities for Spin independent coupling WIMPs with assumptions given in text, limits/ROI from other experiments are from reference 3 - b right ) Simulated electron equivalent energy spectrum from reactor neutrinos induced Argon recoils in spherical detector - see text.}
\label{fig:radish}
\end{figure}

\section{Prospects}

The possibility to operate kg size target mass with low mass nuclei and 100 eV threshold, the performed progress  on background rejection and achieved level of radioactivity lead to anticipate  typical sensitivities for spin independent coupling WIMPs indicated in Figure 4a, with He and Ne nuclei as targets. 
These limits assume a 300 eV nuclear recoil  threshold, a quenching factor deduced from SRIM calculation, a background level of 1 evt for an exposure of 10 kg.d, standard hypothesis  about WIMP velocities and density. 

Another application of such detector is the measurement of coherent neutrino scattering process with neutrinos of energies as the ones from nuclear reactors. Figure 4b shows the simulated expected electron equivalent energy spectrum of Argon recoils induced by neutrinos  from a typical nuclear reactor,  assuming quenching factor from SRIM and instrumental response of a spherical detector. Expected rate in a typical energy search window of 60-260 eV is roughly 20 evts/kg.d/(10$^{13}$ $\nu$/cm$^{2}$/s), that is 20 evts/day for a 1 m diameter sphere filled with Argon at 1 bar, at 10 m from the core of a 0.7 GWt nuclear plant. If the background rate in the same energy region can be made around 100 evts/kg.keV.d, this gives a signal/background of 1. 

So an ON/OFF type experiment can provide a very statistically significant  probe of this process within months of data taking. There are some experimental issues to tackle, in particular the achievable background level near a nuclear reactor and the exact quenching factor values in the energy region of interest. 

A project of a 2 m diameter sphere immersed in a large water shield is ongoing. It would allow to increase the sensitivity to light dark matter search by two orders of magnitude and would also constitute a prototype for a larger sphere for SuperNovae neutrino detection. 

Another astro-particle physics application is the unique capability of such detector to detect  Kaluza-Klein axions~\cite{javier}.

Lastly, this type of detector can also be used for neutron spectroscopy (with nitrogen, patented~\cite{patent}), and developments are ongoing for low level radon and neutron counting and gamma ray spectroscopy in harsh environment.

\section*{Acknowledgments}

The help of the technical staff of the Laboratoire Souterrain de Modane is gratefully acknowledged. The low activity prototype operated in LSM has been partially funded within the European Commission astroparticle program ILIAS (Contract R113-CT-2004- 506222).

\end{document}